\documentclass[conference]{IEEEtran}
\usepackage{color}
\usepackage{url}
\usepackage{float}
\usepackage{array}
\usepackage{xcolor}
\usepackage{amsmath}
\usepackage{amssymb}
\usepackage{balance}
\usepackage{caption}
\usepackage{makecell}
\usepackage{graphicx}
\usepackage{enumitem}
\usepackage{multirow} 
\usepackage[utf8]{inputenc}
\usepackage[export]{adjustbox}
\usepackage[linewidth=0.0pt]{mdframed}
\usepackage[sort&compress,numbers]{natbib}

\makeatletter

\makeatletter
\newcommand*\titleheader[1]{\gdef\@titleheader{#1}}
\AtBeginDocument{%
  \let\st@red@title\@title%
  \def\@title{%
    \vskip-1.4em \bgroup\normalfont\large\centering\@titleheader\par\egroup
    \vskip1.0em\st@red@title}
}
\makeatother

\title{Extractive Summarization of Related Bug-fixing Comments in Support of Bug Repair}

\titleheader{2021 IEEE/ACM International Workshop on Automated Program Repair (APR)}

\begin{document}

\author{
Rrezarta Krasniqi \\
 Dept. of Computer Science and Engineering \\
 University of Notre Dame \\
 Notre Dame, IN, USA \\
 rkrasniq@nd.edu
}

\maketitle

\begin{abstract}
When developers investigate a new bug report, they search for similar previously fixed bug reports and discussion threads attached to them. These discussion threads convey important information about the behavior of the bug including relevant \emph{bug-fixing comments}. Often times, these discussion threads become extensively lengthy due to the severity of the reported bug. This adds another layer of complexity, especially if relevant bug-fixing  comments intermingle with seemingly unrelated  comments. To manually detect these relevant comments among various cross-cutting discussion threads can become a daunting task when dealing with high volume of bug reports. To automate this process, our focus is to initially extract and detect comments in the context of \emph{query relevance}, the use of \emph{positive language}, and \emph{semantic relevance}. Then, we merge these comments in the form of a summary for easy understanding. Specifically, we combine Sentiment Analysis, and the TextRank Model with the baseline Vector Space Model (VSM). Preliminary findings indicate that bug-fixing comments tend to be positive and there exists a semantic relevance with comments from other cross-cutting discussion threads. The results also indicate that our combined approach improves overall ranking performance against the baseline VSM.
\end{abstract}
\section{Introduction}
\label{sec:introduction}
A ``bug fixing'' comment is a comment in a conversation thread attached to a bug report that describes the bug fix. When developers report bugs via issue tracking systems, the conversations between the end-users and developers are presented as a collection of threads. When a fix is communicated, the person who has made the fix or recommends a fix often concludes the conversation with a more-positive language that summarizes the problems and the repairs. Recent studies have indicated that sentiment patterns were noticeable in open-source bug report discussion threads~\cite{maalej2015bug, blaz2016sentiment}. We also observed, that positive sentiment patterns were noticeable when bug reports were concluded as `fixed'. Our broader objective is to automate this process suitably for large-scale open-source platforms. However, there are two emerging factors that need to be taken into account when dealing with open-source platforms~\cite{fitzgerald2006transformation, feller2002understanding,ibrahim2010should}. The most pressing ones span across two directions, primarily \textbf{time dimension} and \textbf{content dimension}. When dealing with the former one, studies showed that open-source platforms were plagued with thousands of bug reports daily~\cite{anvik2005coping, nurolahzade2009role}. Due to this sheer volume of bugs, it becomes prohibitively time consuming for the developers to fix these bugs efficiently and effectively~\cite{marks2011studying}. When dealing with the latter one, a recent qualitative study conducted by Arya \emph{et}. al~\cite{arya2019analysis} showed that some of these issue discussion threads over time become extremely lengthy and complex to comprehend because both developers and end-users carry different backgrounds. Consequently, this communication scenario leads to lengthy discussions. In hindsight, the content factor poses a challenge to developers who must reason which information hidden within these lengthy discussion threads is relevant and useful for finding the root cause for the new bugs. This adds another layer of complexity, especially if relevant bug-fixing comments intermingle with seemingly unrelated comments. Motivated by these observations, this work presents a three-pronged approach that recommends bug-fixing comments extracted from crosscutting issue discussion threads that are beneficial for recommending a bug fix. These bug-fixing comments are analysed and extracted by considering \emph{user query relevance}, \emph{positive language}, and \emph{semantic relevance}. Specifically, we combine Sentiment Analysis (SA), and the TextRank Model (TR) with the baseline Vector Space Model (VSM) as employed by Zhou et \emph{al}.~\cite{zhou2012should}. We use SA to detect positive sentiments in the comments. We then use the TR heuristic to build the semantic connections amongst similar related comments from those cross-cutting discussion threads. \\\vspace{-14pt}
\section{Overview of Approach}
Our approach recommends comments extracted from issue-tracking cross-cutting discussion threads in three steps. First, it uses the baseline VSM to retrieve user query-related comments with initial keyword scores. Second, it uses Sentiment Analysis (SA) to compute sentiment keyword scores based on lookup \emph{bonus and penalty opinion lexicon}, and re-rank comments written in a more positive style from step-1. Finally, it uses meta-heuristic technique such as TextRank (TR) to re-rank keywords based on their co-occurrence and their order of importance from step-2 that are semantically connected. In sum, the above steps represent a combined linear weighted function where keyword weights of each comment are influenced from the combined scores obtained by VSM, SA, and TR.

\subsection{RetroRank Tool}
\label{sec:recommendation_tool}
We have built a GUI-based tool that we refer to as RetroRank. In a nutshell, RetroRank takes as an input \emph{a user query} (e.g., a set of keywords extracted from bug's title or bug's long description) and returns as an output \emph{a list of recommended bug-fixing comments} extracted from cross-cutting discussion threads. RetroRank relies on MySQL database where it stores all OSS bug reports and bug repositories. RetroRank builds upon two existing packages. \emph{Tkinter} is a standard GUI library that is used to build the GUI framework. \emph{Sklearn} is another library that is used to build the underlying functionalities of our approach. Broadly speaking, RetroRank is designed to adapt to any project that supports modern development practices. The inclusion of `pre-merge' mechanism is an example where relevant comments can be used as part of code review~\cite{halloran2002high,mockus2002two}. 
\renewcommand{\arraystretch}{0.5}
\begin{table}[!t]
\caption{Statistical summary of the results. Value \emph{n} denotes the number of bugs evaluated in terms of ranking performance. $\mu$ denotes average ranking position obtained by each configuration. The rest represent Student's Paired t-Test metrics.}
\vspace{-0.25cm}
\label{tab:stat_tests}
\resizebox{\columnwidth}{!}{%
\begin{tabular}{|c|c|c|c|c|c|c|c|}
\hline
Metric & Approach & \emph{n} & $\mu$ & \emph{p} & \emph{t} & \emph{t\_crit} & Decision \\ \hline\hline
Config-1 & \begin{tabular}[c]{@{}c@{}}VSM+SA+TR\\ VSM+SA\end{tabular} & 25 & \begin{tabular}[c]{@{}c@{}}1.8\\ 3.4\end{tabular} & 2.8E-05 & -4.812 & 2.0301 & Reject \\ \hline
Config-2 & \begin{tabular}[c]{@{}c@{}}VSM+SA+TR\\ VSM+TR\end{tabular} & 25 & \begin{tabular}[c]{@{}c@{}}1.8\\ 3.7\end{tabular} & 2.9E-06 & -5.557 & 2.0301 & Reject \\ \hline
Config-3 & \begin{tabular}[c]{@{}c@{}}VSM\\ VSM+TR\end{tabular} & 25 & \begin{tabular}[c]{@{}c@{}}9.1\\ 3.7\end{tabular} & 1.3E-09 & 8.146 & 2.0638 & Reject \\ \hline
Config-4 & \begin{tabular}[c]{@{}c@{}}VSM\\ VSM+SA\end{tabular} & 25 & \begin{tabular}[c]{@{}c@{}}9.1\\ 3.4\end{tabular} & 2.0E-09 & 8.004 & 2.0638 & Reject \\ \hline
Config-5 &  \begin{tabular}[c]{@{}c@{}}VSM\\ VSM+SA+TR\end{tabular} & 25 & \begin{tabular}[c]{@{}c@{}}\textbf{9.1}\\ \textbf{1.8}\end{tabular} & \textbf{3.0E-11} & 9.513 & 2.0638 & Reject \\ \hline
\end{tabular}%
}
\vspace{-0.50cm}
\end{table}

\vspace{-0.40cm}
\section{Evaluation Strategy}
\label{sec:evaluation}
\subsection{Preliminary Synthetic Study}
\label{sec:synthetic_study}
In the synthetic study, we analyzed $25$ real bugs to compare our approach against several configurations. Our evaluation strategy employed four steps. First, we manually identified the goldset of bugs and bug-fixing comments. Second, we recruited 12 developers to create `search queries' for new bugs. Third, we used those `user created queries" from the previous step to run each configuration. Finally, based on the results retrieved from each configuration, we counted and marked the ranking position where the goldset (i.e., bug-fixing comments from past resolved bugs) appeared in each configuration's retrieved list. Due to space limitation, we only reported the significance of ranking performed of each configuration as denoted in Table~\ref{tab:stat_tests}, where $\mu$ denotes the average ranking obtained by each configuration. We plan to extend this work by fully reporting its findings.

\subsection{Preliminary Results: Overall Ranking Performance}
\label{sec:results}
Preliminary results indicate that VSM+SA+TR achieved the highest performance, followed by VSM+SA, VSM+TR, and the baseline VSM as employed by Zhou et. al~\cite{zhou2012should}. Findings in Table~\ref{tab:stat_tests} showed that the average ranking position $\mu$ of the recommended comment for VSM+SA+TR was $1.8$, whereas for the VSM was $9.1$. This indicates, VSM+SA+TR placed the recommended comment on average around first position, versus VSM which placed it around the ninth position. We also noted that both VSM+SA and VSM+TR ranked in a higher position than the baseline VSM. Overall, results showed that differences in ranking position were statistically significant as reported in Table~\ref{tab:stat_tests}, H$_5$ ($3.0$E$-11$) with p-value of $0.05$ and a confidence level of $0.95$. Student’s Paired t–Test~\cite{smucker2007comparison} was used as a primary metric as done in other studies~\cite{GouesDFW12,WangLJLL11,bettenburg2008duplicate}.
\section{Conclusion and Evaluation Plans}
\label{sec:conclusion}
In this position paper, we presented a novel idea of how issue tracking cross-cutting discussion threads can be used as the core content in support of bug repair. As part of the evaluation plan, we will conduct a larger user study where we will empirically compare the best-performing approach to the baseline VSM. Our objective is to also measure comment(s) \emph{relevance} returned by each approach. Additionally, we will perform a qualitative user-study where the rationale behind our work will be explained in-depth.

\bibliographystyle{abbrv}
\balance \footnotesize
\bibliography{main}

\end{document}